\documentclass[aps, floatfix, prd, twocolumn, 10pt, superscriptaddress, nofootinbib]{revtex4-1}
\usepackage{amsmath, amsthm, amssymb}
\usepackage{bbm} 
\usepackage{epsfig}
\usepackage{slashed}
\usepackage{color}

\graphicspath{{graphics/}}
\usepackage{hyperref}

\renewcommand{\=}{\;=\;}
\newcommand{\+}{\;+\;}

\newcommand{\eq}[1]{Eq.~(\ref{#1})}

\newcommand{\fig}[1]{Fig.~\ref{#1}}

\newcommand{\Tr}{\mbox{Tr }}
\newcommand{\SNG}{\mathcal{S}}

\newcommand{\sumint}{\sum\hspace{-4.5mm}\int}
\newcommand{\va}[1]{\vert\vec #1 \vert}

\newcommand{\bit}{\begin{itemize}}
\newcommand{\eit}{\end{itemize}}
\newcommand{\beq}{\begin{equation}}
\newcommand{\eeq}{\end{equation}}
\newcommand{\bea}{\begin{align}}
\newcommand{\eea}{\end{align}\]}
\newcommand{\lb}{\left(}
\newcommand{\rb}{\right)}
\def\bseq#1\eseq{\begin{equation}\begin{split}#1\end{split}\end{equation}}

\usepackage{color}
\usepackage{ulem}

\begin{document}
 \title{Dyson-Schwinger Approach to Color-Superconductivity:\\
 Effects of Selfconsistent Gluon Dressing}
 \date{\today}
 \author{D. M\"uller} 
 \affiliation{Theoriezentrum, Institut f\"ur Kernphysik, Technische Universit\"at Darmstadt, Germany}
  
 \author{M. Buballa}
 \affiliation{Theoriezentrum, Institut f\"ur Kernphysik, Technische Universit\"at Darmstadt, Germany}

 \author{J. Wambach}
 \affiliation{Theoriezentrum, Institut f\"ur Kernphysik, Technische Universit\"at Darmstadt, Germany}
 \affiliation{ECT*, Villazzano (TN), Italy}

\begin{abstract}
The phase diagram of dense QCD at nonvanishing temperatures and large quark chemical potentials is studied with Dyson-Schwinger equations for $2+1$ quark flavors, focusing on color-superconducting phases with 2SC and CFL-like pairing. The truncation scheme of our previous investigations~\cite{Muller:2013pya} is extended to include the dressing of gluons with selfconsistently determined quarks, i.e.,  taking into account the dynamical masses and superconducting gaps of the quarks in the gluon polarization. As a consequence the gluon screening is reduced, leading to an enhancement of the critical temperatures of the color-superconducting phases by about a factor of 2 as compared to the case where the gluons are dressed with bare quarks. We also calculate the Debye and Meissner masses of the gluons and show that they are consistent with weak-coupling results.
\end{abstract}

\maketitle

\section{Introduction}

The phase structure of QCD at high but not asymptotically high baryon densities remains an unresolved problem.
While in recent years, thanks to substantial effort in lattice gauge theory, consensus has been reached about the equation of state at finite temperature but vanishing chemical potential~\cite{Borsanyi:2010bp,Bazavov:2014pvz}, 
the application of these techniques to large chemical potentials is inhibited by the fermion sign problem.
In this context functional methods, such as the Functional Renormalization Group (FRG)~\cite{Braun:2008pi,Braun:2009gm,Herbst:2013ail, Herbst:2013ufa} or Dyson-Schwinger equations (DSEs)~\cite{Muller:2013pya,Fischer:2011mz,Fischer:2012vc,Fischer:2013eca,Fischer:2014ata,Eichmann:2015kfa,Muller:2013tya,Nickel:2006vf,Nickel:2006kc,Nickel:2008ef}, have become important alternative tools. They successfully reproduce the lattice-QCD results at zero density but in addition also provide access to the finite-density regime.

Using DSEs, various studies of the QCD phase diagram at nonzero temperature and densities have been 
performed, mostly focusing on  low and moderate chemical potentials. These studies confirm the lattice-result of a crossover at low densities and predict a first-order transition at higher densities\footnote{Allowing for non-uniform structures, there can also be an inhomogeneous phase in this regime~\cite{Muller:2013tya}.} 
with a critical endpoint at quark chemical potentials typically around $\mu = 150-200$~MeV~\cite{Fischer:2012vc,Fischer:2013eca,Fischer:2014ata,Eichmann:2015kfa}. 

Some time ago, DSEs have also been applied to QCD at zero temperature and higher chemical potential 
to study color superconductivity, i.e., the formation of quark Cooper pairs~\cite{Nickel:2006vf,Nickel:2006kc,Nickel:2008ef}. Both, the two-flavor superconducting (2SC) and the color-flavor locked (CFL) pairing patterns have been investigated. In our previous work~\cite{Muller:2013pya} we extended the analysis to nonvanishing temperature, considering $2+1$ flavors with massless up- and down quarks and a realistic strange quark mass. We found that the CFL phase is favored at low temperature and chemical potentials larger than $500-600$~MeV, while at lower chemical potentials there is a 2SC phase, which also extends to higher chemical potentials in a narrow temperature band above the CFL phase. The critical temperatures to the normal conducting phase are of the order of $20-30$~MeV.

However, these results should not be considered as final. Although, in principle, DSEs provide an exact approach to QCD, they form an infinite set of integro-differential equations, which needs to be truncated in order to be solved in practice. In Ref.~\cite{Muller:2013pya} a truncation was employed where the quark loop of the gluon polarization function was evaluated in hard-thermal-loop-hard-dense-loop (HTL-HDL) approximation. 
This implies that the vacuum parts of the quark loops were neglected
while the medium contrubutions were evaluated with bare quarks instead of dressed ones. 
This has the advantage that the quark loop can be calculated analytically, keeping the numerical effort similar to a quenched calculation where quark effects on the gluon propagator are neglected completely.

On the other hand, while this approximation is reasonable in the high-temperature regime, where the superconducting gap vanishes and the dynamical quark masses are small, it becomes obviously questionable in color superconducting phases or phases with broken chiral symmetry, including the vacuum. Indeed it was found in Ref.~\cite{Fischer:2012vc} that including the dynamical quark mass in the gluon polarization leads to a sizeable shift of the chiral critical endpoint relative to the HTL-HDL result~\cite{Fischer:2011mz}. Moreover, in Ref.~\cite{Muller:2013pya} we were not able to match vacuum and in-medium observables simultaneously: 
Adjusting the parameters to reproduce the lattice value of the chiral critical temperature, the value of the pion decay constant $f_\pi$ in vacuum was overestimated by $40\%$, while a fit to $f_\pi$ yields a critical temperature below 100~MeV.

Here we therefore want to present an improved version of our analysis in Ref.~\cite{Muller:2013pya}, using a truncation where the gluon polarization is calculated selfconsistently with dressed quarks. While for the critical endpoint and the crossover region at low chemical potentials the effects of this selfconsistent treatment have already been studied in Ref.~\cite{Fischer:2012vc},  the novel aspect of the present work is the extension to color superconducting phases, in particular taking into account the anomalous selfenergy of the quark propagator.
As an additional result, we also obtain the Debye and Meissner masses of the gluons. Evaluating them in the weak-coupling limit, they provide an additional consistency check for our truncation scheme, in particular for the quark-gluon vertex.  

The remainder part of this paper is organized as follows. In Section \ref{sec.dse} we briefly recap the basic Dyson-Schwinger formalism of our previous work. In Section \ref{sec.trunc} we introduce the improved truncation scheme, taking special care about a proper renormalization of the gluon polarization loop and a consistent treatment of the quark-gluon vertex function. The results for the color-superconducting condensates and phase diagrams are presented in Section \ref{sec.results}. In Section \ref{sec.debye} we discuss the Debye and Meissner masses of the gluons, both, in the weak-coupling limit and at strong coupling. Finally, in Section \ref{sec:conclusions}, we draw conclusions.

\section{Dyson-Schwinger formalism for color-superconducting phases}
\label{sec.dse}

In this section we briefly summarize the Dyson-Schwinger formalism of our previous work on color superconducting matter~\cite{Muller:2013pya}. We restrict ourselves to those aspects which are relevant for our present studies and refer to that reference for further technical details.

\begin{figure}[h]
	\centering
		\includegraphics[scale=0.8]{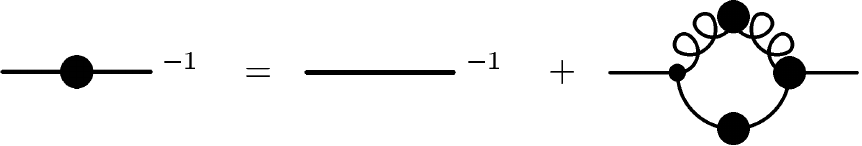}	
	\caption{DSE of the quark propagator. Plain and curly lines represent quark and gluon propagators, respectively, while thick dots represent dressed quantities.}
\label{fig:qdse}
\end{figure}

We work in Landau gauge and solve the Dyson-Schwinger equation for the dressed quark propagator, depicted in \fig{fig:qdse} in imaginary time. 
At temperature $T$ and chemical potential $\mu$ the equation is given by
\beq
\label{eq:qdse}
\SNG^{-1}(p) \= Z_2(\SNG^{-1}_0(p) + \Sigma(p))\,,
\eeq
with $p=(p_4,\vec p)=(\omega_n+i\mu,\vec p)$ and the fermionic Matsubara frequency $\omega_n=(2n+1)\pi T$.
$Z_2$ is the quark wave-function renormalization constant. As we study color superconductivity, the bare and dressed quark propagators $\SNG_0$ and $\SNG$ as well as the quark self-energy $\Sigma$ are $2\times 2$ matrices in Nambu-Gorkov (NG) space
\beq
 \SNG_0(p) \= \begin{pmatrix} S_0^+(p) & 0\\ 0 & S_0^-(p) \end{pmatrix},
\eeq
\beq
  \SNG(p) \= \begin{pmatrix} S^+(p) & T^-(p) \\ T^+(p)  & S^-(p) \end{pmatrix},
\eeq
\beq
   \Sigma(p) \= \begin{pmatrix} \Sigma^+(p) & \Phi^-(p) \\ \Phi^+(p)  & \Sigma^-(p) \end{pmatrix}.
\eeq
The normal components $S^\pm$ and $\Sigma^\pm$ correspond to particle and charge conjugate particle propagators and self-energies, while the off-diagonal components $T^\pm$ and $\Phi^\pm$ are related to the color-superconducting condensates. In addition, each component has a Dirac, color and flavor substructure, where we consider $N_c=3$ colors and $2+1$ flavors, unless noted otherwise. The color-flavor structure of the dressed propagator is then parametrized by
\bseq\label{eq:SPTM}
{S^+}(p)&
\= \sum_i {S^+_i}(p)\,P_i,
\\
T^+(p) &
\= \sum_i T^+_i(p)\,M_i,
\eseq
with the matrices $P_i$ and $M_i$ defined in Appendix \ref{app:par}. The components have the following tensor structure in Dirac space:
\bseq\label{eq:prop_dirac}
&{S_i^+}^{-1}(p) \= \\& -i(\omega_n+i\mu)\gamma_4 C_i^+(p) -i\slashed{\vec p}A_i^+(p)+B_i^+(p) -i \gamma_4\frac{\slashed{\vec p}}{\va p} D_i^+(p),\\
&T_i^+(p) \= \\&\left(\gamma_4\frac{\slashed{\vec p}}{\va p}T_{A,i}^+(p) + \gamma_4T_{B,i}^+(p) + T_{C,i}^+(p)+\frac{\slashed{\vec p}}{\va p}T_{D,i}^+(p)\right)\gamma_5\,,
\eseq
each with 4 scalar dressing functions. The self-energies $\Sigma^+(p)$ and $\Phi^+(p)$ can be decomposed analogously, while the remaining NG components ($S^-$, $T^-$, $\Sigma^-$, $\Phi^-$) can be obtained from the $+$ components by using symmetry relations~\cite{Rischke:2003mt,Nickel:2006vf}.

The bare propagators are diagonal in color and flavor, and are given by
\beq\label{eq:prop_bare}
\lb S^{\pm}_0(p)\rb^{-1} = -i\gamma_4\omega_n \pm\gamma_4 \mu  -i\slashed{\vec p} + Z_m m_f
\eeq
with the mass renormalization constant $Z_m$ and the flavor-dependent bare quark mass $m_f$. The renormalization constants are fixed by the renormalization condition in vacuum
\beq
\lb S^{+}(p^2=\nu^2)\rb^{-1}=\lb -i\slashed p +m_f\rb\vert_{p^2=\nu^2}
\eeq
at an arbitrary renormalization point $\nu$.

Introducing the abbreviation 
${\sum\hspace{-3.5mm}\int\hspace{1mm}}_q \equiv T\sum\int\frac{d^3q}{(2\pi)^3}$, 
where the summation is over the Matsubara frequencies, the quark self-energy reads
\beq\label{eq:qse}
Z_2\Sigma(p) \= g^2\sumint\limits_q\,\Gamma_\mu^{a,0}\SNG(q)\Gamma^b_\nu(p,q)D^{ab}_{\mu\nu}(k=p-q)\,,
\eeq
with the QCD coupling constant $g$, the bare and the dressed quark-gluon vertices,
$\Gamma_\mu^{a,0}$ and $\Gamma_\nu^b$ and the dressed gluon propagator $D^{ab}_{\mu\nu}$, respectively.

The NG structure of the dressed vertex is generally parametrized as
\beq\label{eq:vertex_full}
  \Gamma^a_{\mu}(p,q) \= \begin{pmatrix} \Gamma^{a,+}_\mu(p,q) & \Delta^{a,-}_\mu(p,q)\\ \Delta^{a,+}_\mu(p,q) &  \Gamma^{a,-}_\mu(p,q) \end{pmatrix}\,,
\eeq
with momentum-dependent diagonal and off-diagonal components, which will be specified in Sec.~\ref{sec.trunc.qgv}. The bare vertex is momentum and flavor independent and diagonal in NG space.
It is given by 
\beq\label{eq:Gamma0}
\Gamma^{a,0}_{\mu} \= Z_{1F}
\begin{pmatrix} \gamma_\mu\frac{\lambda_a}{2} & 0
\\ 
0 & -\gamma_\mu\frac{\lambda_a^{T}}{2} \end{pmatrix}
\equiv Z_{1F} \gamma_\mu \frac{\Lambda_a}{2}\,,
\eeq
where $\lambda^a$ are the 8 Gell-Mann matrices in color space and $Z_{1F}$ denotes the renormalization constant of the quark-gluon vertex.

Finally we parametrize the dressed gluon propagator as
\beq\label{eq:Dfull}
D^{ab}_{\mu\nu}(k) \= \frac{Z^{ab}_{TT}(k)}{k^2} P^T_{\mu\nu}(k) + \frac{Z^{ab}_{TL}(k)}{k^2} P^L_{\mu\nu}(k),
\eeq
exploiting the fact that it is transverse in 4-dimensional space. To that end we define the 4-dimensional transverse and longitudinal projectors
\bseq
T_{\mu\nu}&=\delta_{\mu\nu}-\frac{k_\mu k_\nu}{k^2}\\
L_{\mu\nu}&=\frac{k_\mu k_\nu}{k^2}
\eseq
as well as their 3-dimensional counterparts
\begin{alignat}{1}
P^T_{ij}(k)&=\delta_{ij}-\frac{k_i k_j}{\vec k^2}, \quad
P^T_{44}(k)=P^T_{i4}(k)=P^T_{4i}(k)= 0\,,
\nonumber\\
P^L_{\mu\nu}(k)&=T_{\mu\nu}(k)-P^T_{\mu\nu}(k)\,,
\end{alignat}
where greek indices run from 1 to 4, and latin indices from 1 to 3. The corresponding dressing functions of the gluon propagator, $Z_{TT,}^{ab}(k)$ and $Z_{TL}^{ab}(k)$, will be specified in Sec.~\ref{sec.trunc.glu}.

\section{Truncation}
\label{sec.trunc}

The quark DSE, \eq{eq:qdse}, is an exact equation and can be solved once the dressed gluon propagator and quark-gluon vertex are known.  These are given as solutions of their own DSEs, depending on higher $n$-point functions. Therefore truncations are necessary in order to get a closed set of equations. In this section, we describe our improved truncation scheme, extending our previous work~\cite{Muller:2013pya}.

\subsection{Truncation of the gluon propagator}
\label{sec.trunc.glu}

\begin{figure}
	\centering
		\includegraphics[scale=0.8]{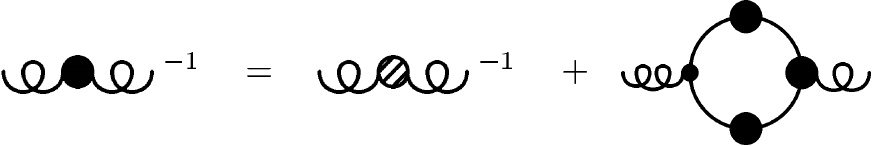}	
	\caption{Truncated gluon DSE. Plain and curly lines represent quark and gluon propagators, respectively, while thick dots represent dressed quantities. The propagator with the shaded dot is the dressed Yang-Mills gluon propagator.}
\label{fig:gdse}
\end{figure}

The DSE for the dressed gluon propagator consists of a pure Yang-Mills part and a quark-loop diagram, as shown in \fig{fig:gdse}. 
Similar to other works~\cite{Aguilar:2012rz,Fischer:2012vc}, we neglect the quark back-coupling
on the Yang-Mills sector. Instead, the quark loop is added perturbatively to the equation,
without modifying the Yang-Mills propagator. This has the advantage that the Yang-Mills diagrams do not need to be calculated explicitly and we can use lattice data. The truncated gluon DSE in this approximation is then given by
\beq\label{eq:gdse}
  D^{-1,ab}_{\mu\nu}(k) = D^{-1,ab}_{\mu\nu, YM}(k) + \Pi^{ab}_{\mu\nu}(k),
\eeq
with the Yang-Mills gluon propagator
\beq\label{eq:DYM}
  D^{ab}_{\mu\nu, YM}(k) \= \frac{Z_{TT}^{YM}(k)}{k^2} P^T_{\mu\nu}(k)\+\frac{Z_{TL}^{YM}(k)}{k^2} P^L_{\mu\nu}(k)
\eeq
and the Yang-Mills dressing functions $Z_{TT,TL}^{YM}(k)$, obtained from fits to lattice data \cite{Fischer:2010fx}. Quark effects on the gluon are taken into account in the gluon polarization 
\beq\label{eq:gpol}
  \Pi^{ab}_{\mu\nu}(k) \= -\frac{g^2}{2}\sumint\limits_q\Tr \left(\Gamma_{\mu}^{a,0}\SNG(p)\Gamma_{\nu}^{b}(p,q)\SNG(q) \right),
\eeq
where $p=k+q$.

This truncation is self-consistent on the level of quarks, as the polarization loop depends on dressed quark propagators. In our previous study~\cite{Muller:2013pya}, we dropped the self-consistency at this point and evaluated the quark loop in a hard-thermal-loop--hard-dense-loop approximation (HTL-HDL) with bare quark propagators. The use of bare quark propagators in the gluon DSE makes it independent of the dressed quark propagator and the numerical effort is similar to a pure rainbow truncation. On the other hand, it neglects important contributions to the gluon polarization, such as those of quark masses and color-superconducting condensates.  This can be tolerated in the chirally restored phase at high temperatures, where these contributions are small, but becomes questionable in the chirally broken phase and in the color-superconducting regime. Here, we therefore calculate the quark loop in the polarization function \eq{eq:gpol} with self-consistent fully dressed Nambu-Gor'kov quark propagators.

Although the full gluon propagator must be transverse, ensured by Slavnov-Taylor identities in full QCD, the polarization loop, \eq{eq:gpol}, may have longitudinal contributions. In the full gluon DSE these would be canceled by longitudinal contributions of Yang-Mills diagrams. However, as we do not calculate the quark back-coupling to the Yang-Mills system, we do not get this cancellation. Therefore we drop the longitudinal part of the polarization loop by hand. 

Another technical complication arises from the fact that for the numerical evaluation of the polarization loop we have to introduce a momentum cut-off. This generates artificial quadratic divergencies $\sim \Lambda^2$, which have to be regularized. These divergencies do not occur in dimensional regularization, but this scheme cannot be employed in the numerical calculations.\footnote{We checked, however, that for the limiting case of bare propagators in vacuum, our scheme leads to the same result as for dimensional regularization.}
In vacuum the regularization can be done easily by subtracting a constant term $\Pi_{\mu\nu}^{ab}(0)$. In the medium one has to be more careful to preserve the thermal and dense gluon Debye masses 
$\sim T^2, \mu^2$ and to ensure that they agree with weak coupling results. We therefore use a scheme similar to the one proposed by Brown and Pennington~\cite{Brown:1988bm} which defines the regularized vacuum polarization $\Pi_{reg}^{BP}(k)$ by
\beq
\Pi_{reg}^{BP}(k) = \frac{1}{3}\left(\delta_{\mu\nu}-4\frac{k_\mu k_\nu}{k^2}\right)\Pi_{\mu\nu}(k).
\eeq
Decomposing $\Pi_{\mu\nu}$ into a transverse and a (4-dimensional) longitudinal part,
\beq
\Pi_{\mu\nu}(k) = T_{\mu\nu} \Pi_T(k) + L_{\mu\nu} \Pi_L(k),
\eeq
the regularized gluon polarization is given by
\beq
 \Pi_{reg}^{BP}(k) = \Pi_T(k) - \Pi_L(k).
\eeq
Alternatively, we can subtract $\Pi_L(0)$ instead of $\Pi_L(k)$. This only removes a constant contribution $\sim \Lambda^2$, i.e., only the divergent contribution. After renormalization as defined below, both subtraction schemes give identical results for bare quarks. However, for numerical reasons, we prefer to subtract $\Pi_L(0)$.
Extending this scheme to the medium, the regularized polarizations are thus given by
\bseq\label{eq:bp_reg}
 \Pi_{TT,reg}(k) &= \Pi_{TT}(k) - \Pi_L(0)\\
 \Pi_{TL,reg}(k) &= \Pi_{TL}(k) - \Pi_L(0)~.
\eseq
Evaluated with massless bare propagators, \eq{eq:prop_bare}, these expressions reproduce the HTL-HDL results.

The remaining logarithmic divergencies are renormalized by
\beq\label{eq:gluren}
\Pi_{i,ren}(k) = \Pi_{i,reg}(k) - \frac{k^2}{\nu^2}\Pi_{i,reg}(\nu)
\eeq
for both components $i=TT,TL$. Although, in principle, the subtraction term should be evaluated in vacuum, in order to achieve numerical convergence we must evaluate both polarization functions at the same temperature and chemical potential, so that the integrands are given at the same momentum mesh. However, the difference is negligible if the renormalization point is large enough.We choose the renormalization point at
$|\nu| =7.65$ GeV, which is the scale where the vacuum Yang-Mills gluon dressing is $Z^{YM}(\nu) = 1$. 
For numerical stability, it is useful to choose the direction of the 4-vector $\nu$ to be equal to the direction of $k$. 

Inserting the renormalized polarization tensor into \eq{eq:gdse}, we get the dressed gluon propagator
\bseq\label{eq:Ddressed}
  D^{ab}_{\mu\nu}(k) &\= \frac{Z_{TT}^{YM}(k)}{k^2 + Z_{TT}^{YM}(k)\Pi^{ab}_{TT,ren}(k)} P^T_{\mu\nu}(k)\\ &\+\frac{Z_{TL}^{YM}(k)}{k^2 + Z_{TL}^{YM}(k)\Pi^{ab}_{TL,ren}(k)} P^L_{\mu\nu}(k).
\eseq
Some details on the numerical evaluation are given in Appendix \ref{app:num_calc}.

\subsection{Truncation of the quark-gluon vertex}
\label{sec.trunc.qgv}

The second quantity which needs to be truncated is the quark-gluon vertex, which enters the calculation at two different places: in the quark self-energy $\Sigma$, \eq{eq:qse}, and in the gluon polarization tensor $\Pi$, \eq{eq:gpol}. In principle, the same vertex function should be used in both cases, but,  as we will explain, we have to make slightly different choices. 

In both cases we take an Abelian part, 
\beq\label{eq:vertex_abel}
\Gamma^a_{\mu,abel}(p,q) \= \gamma_\mu \frac{\Lambda_a}{2}\Gamma(p,q),
\eeq
which is a straight-forward generalization of the bare vertex defined in  \eq{eq:Gamma0}. $\Gamma(p,q)$  is a dressing function, which is modeled to have the correct perturbative running in the ultraviolet and an infrared enhancement, as necessary to generate chiral symmetry breaking  in vacuum. In general it depends on both quark momenta, $p$ and $q$,  but, as in~\cite{Muller:2013pya}, we use the ansatz $\Gamma(p,q) = \Gamma(\kappa^2)$ with
 \beq\label{eq:vertex1}
\frac{\Gamma(\kappa^2)}{Z_2\tilde Z_3} = \frac{d_1}{d_2+\kappa^2} + \frac{\kappa^2}{\kappa^2+\Lambda^2}\left(\frac{\beta_0\alpha(\nu)\ln(\kappa^2/\Lambda^2+1)}{4\pi}\right)^{2\delta}\,,
\eeq
depending only on a single momentum variable $\kappa^2$. 

For the quark self-energy, we identify $\kappa$ with the gluon momentum $k = p-q$, i.e.,  in \eq{eq:qse} we take $\Gamma(p,q) = \Gamma(k^2)$, as also done in~\cite{Muller:2013pya}. In the gluon polarization \eq{eq:gpol}, on the other hand, we need to take an explicit dependence on quark momenta to preserve renormalizability.\footnote{For $\Gamma(p,q)=\Gamma(k^2)$, the vertex does not depend on the integration variable in \eq{eq:gpol} and can be factored out of the integral. As a consequence, the renormalization condition \eq{eq:gluren} does not give a finite result anymore. Therefore, the vertex dressing needs an explicit dependence on the quark momenta.}
This could be realized by taking $\kappa^2 = p^2 + q^2$. However, since $p_4$ and $q_4$, as defined below \eq{eq:qdse}, contain a term $i\mu$, the vertex function would then become complex and even have singularities
which are not associated with a physical process. In order to avoid these unwanted effects but keep the renormalizability, in the polarization functions we therefore take $\Gamma(p,q)=\Gamma(\tilde p^2+\tilde q^2)$, 
where $\tilde p$ and $\tilde q$, correspond to $p$ and $q$ at $\mu = 0$, i.e., we drop the $i\mu$ terms. Minor artifacts related to this choice will be discussed in in Sec.~\ref{silverblaze}.

For the parameters in  \eq{eq:vertex1} we adopt the values of Ref.~\cite{Muller:2013pya}. 
The only exception is the parameter $d_1$, 
which is refitted to obtain a critical temperature of $T_c=150$~MeV for the chiral phase transition at $\mu=0$. 
Recalling that in HTL-HDL approximation the vacuum contribution of the gluon polarization by the quarks is neglected,
the overall screening in the selfconsistent scheme is stronger, so that more IR enhancement is needed in 
order to compensate for it.
We find $d_1=14$~GeV$^2$, which is about 50\% higher than the value in~\cite{Muller:2013pya}.

\begin{figure}
	\centering
		\includegraphics[]{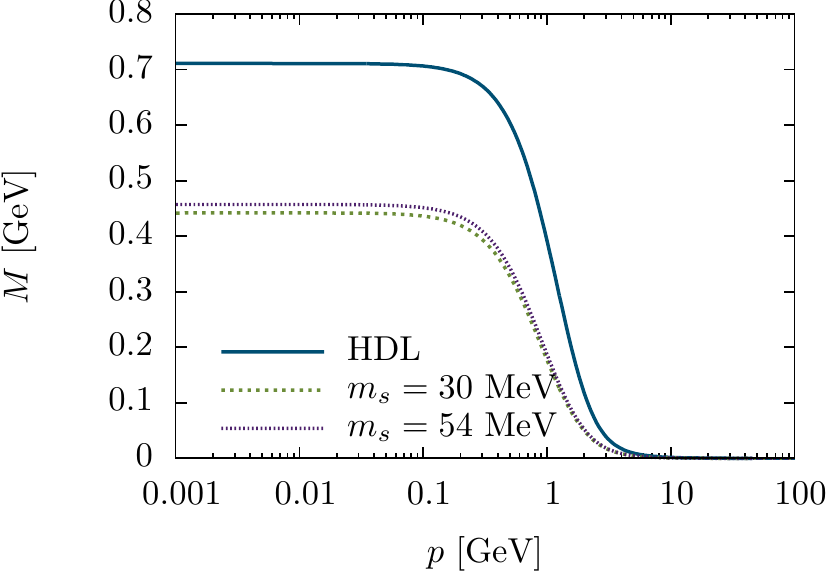}	
	\caption{Vacuum quark mass function $M(p)= B_1(p)/C_1(p)$
	 for chiral up and down quarks in HTL-HDL approximation and with the improved gluon propagator for two different values of the bare strange-quark mass.}
\label{fig:lqmass}
\end{figure}
The corresponding mass functions for up and down quarks in vacuum  are shown in \fig{fig:lqmass} in comparison with the result of our previous work in HDL-HTL approximation~\cite{Muller:2013pya}. It is clearly visible that the improved truncation results in lower quark masses. In particular the values at low momenta are closer to typical constituent quark masses in phenomenological models. Note that, because of the strange-quark contribution to the gluon polarization, the up- and down-quark mass functions depend slightly on the bare strange-quark mass,
which was not the case in HDL-HTL approximation.

For the pion decay constant, which can be calculated via the Pagels-Stokar formula~\cite{Pagels:1979hd}, we find $f_\pi= 95$~MeV. This is much closer to the empirical value of $92.4$~MeV than the HTL-HDL result of our previous work, where we obtained $127$~MeV.

In general, the quark-gluon vertex is constrained by a Slavnov-Taylor identity (STI), which connects it with the quark propagator, the ghost propagator $G(k)$ and the ghost-quark scattering kernel $H^a(p,q)$~\cite{Marciano:1977su},
\beq
-i G^{-1}(k)k_\mu\Gamma^a_{\mu}(p,q) = \SNG^{-1}(p)H^a(p,q)-H^a(p,q)\SNG^{-1}(q)~.
\eeq
As we do not solve the Yang-Mills system explicitly, we have no information about the ghost sector, so that the STI cannot be implemented exactly.
However, for color-superconducting phases, the STI protects important symmetries which are crucial to reproduce the weak-coupling limits for the Debye and Meissner masses, calculated in Refs.~\cite{Rischke:2000qz,Rischke:2000ra}. Therefore we can use the weak-coupling calculations as a guide how to implement the STI approximately. 

If we assume $H^a(p,q) = g(p,q)\frac{\Lambda^a}{2}$ with a scalar function $g(p,q)$, the STI reduces to a restriction similar to the Ward-Takahashi identity (WTI) in QED, except for a multiplicative scalar function $f(p,q)$
\beq\label{eq:bccond}
-i k_\mu\Gamma_{\mu}^{a}(p,q) \propto \SNG^{-1}(p)\frac{\Lambda^a}{2}-\frac{\Lambda^a}{2}\SNG^{-1}(q).
\eeq
This identity can be used to give constraints to the quark-gluon vertex as proposed by Ball and Chiu \cite{Ball:1980ay}. The Ball-Chiu vertex ensures a transverse quark contribution to the gluon polarization,
$k_\mu k_\nu \Pi^{ab}_{\mu\nu}(k) \= 0$.
This is a strict requirement in QED, while in QCD longitudinal parts are in gereral allowed, as discussed before. 
Therefore, we require \eq{eq:bccond} only to be fulfilled in the weak-coupling limit. 
In that case the quark propagators have a simple color-superconducting self-energy
\beq
\label{eq:cssimple}
\Phi^+(p) = \phi_{i}(p)\gamma_5 M_{i},
\eeq
with matrices $M_i=M_{2SC}$ for the 2SC phase and $M_i=M_{sing/oct}$ for the CFL phase, which are defined in Appendix \ref{app:par}.
Using this self-energy in \eq{eq:bccond} and constructing the Ball-Chiu vertex, we obtain
\beq
\Gamma_{\mu}^{a}(p,q) \propto \gamma_\mu\frac{\Lambda^a}{2} + \Gamma_{\mu,CSC}^{a}
\eeq
with
\beq\label{eq:bc_vertex}
\Gamma_{\mu,CSC}^{a} =\frac{ik_\mu}{2k^2}\begin{pmatrix}0 & -(\lambda^a \Phi^- + \Phi^-\lambda^{a,T})\\ (\lambda^{a,T} \Phi^+ +\Phi^+\lambda^a) & 0\end{pmatrix},
\eeq
where $\Phi \equiv (\Phi(p)+\Phi(q))/2$. Thereby we have dropped contributions which depend on derivatives of $\Phi$.

Generalizing this result to the quark propagators in the DSE
we therefore use the vertex
\beq
\Gamma_{\mu}^{a}(p,q) = \lb\gamma_\mu\frac{\Lambda^a}{2} + \Gamma_{\mu,CSC}^{a}\rb\Gamma(p,q)
\eeq
with the full anomalous self-energies $\Phi^\pm$. For normal conducting phases, this reduces to the Abelian vertex \eq{eq:vertex_abel}. Moreover, since $\Gamma_{\mu,CSC}^{a}$ is purely longitudinal, it does not contribute to the quark self-energy,  \eq{eq:qse}, where it is attached to the transverse gluon propagator. In the gluon polarization, on the other hand, it enters through the longitudinal subtraction terms in \eq{eq:bp_reg} and is therefore crucial for a consistent subtraction in color superconducting phases. 
 
\section{Phase structure}\label{sec.results}

After specifying the truncation, the coupled quark and gluon DSEs form a closed system and can be solved numerically. As in our previous publication, we consider chiral up and down quarks and two different values of the strange-quark mass,  $m_s(\nu)=30$~MeV and $m_s(\nu)=54$~MeV, at a renormalization scale of $\nu=100$~GeV. This mass range can be motivated by a perturbative evolution of the PDG values~\cite{Beringer:1900zz} to that scale~\cite{Muller:2013pya}.

The (light-)quark condensate is given by
\beq\label{eq:qq_cond}
\langle \bar q q \rangle = -Z_m Z_2 \sumint\limits_q \Tr_{D,c}(S_{u}^+(q)),
\eeq
where $S_u^+$ denotes the up-quark component of $S^+$. For the color superconducting condensates we define
\beq\label{eq:csc_cond}
{\cal C}_i
\equiv
\langle \psi^T C\gamma_5 \mathcal{O}_i \psi \rangle 
= -Z_2 \sumint\limits_q \Tr(\gamma_5\mathcal{O}_i T^-(q)),
\eeq
with an operator $\mathcal{O}_i$ projecting on the desired component. Throughout this work we restrict the discussion to a condensate only containing up and down quarks,
\beq
    \mathcal{O}_{ud} =  \frac{1}{4}(M_1 - M_2),
\eeq
and a condensate which also involves strange quarks,
\beq
    \mathcal{O}_{uds} = \frac{1}{8}(M_6 + M_7 - M_4 -M_5).
\eeq

\begin{figure}
	\centering
		\includegraphics[]{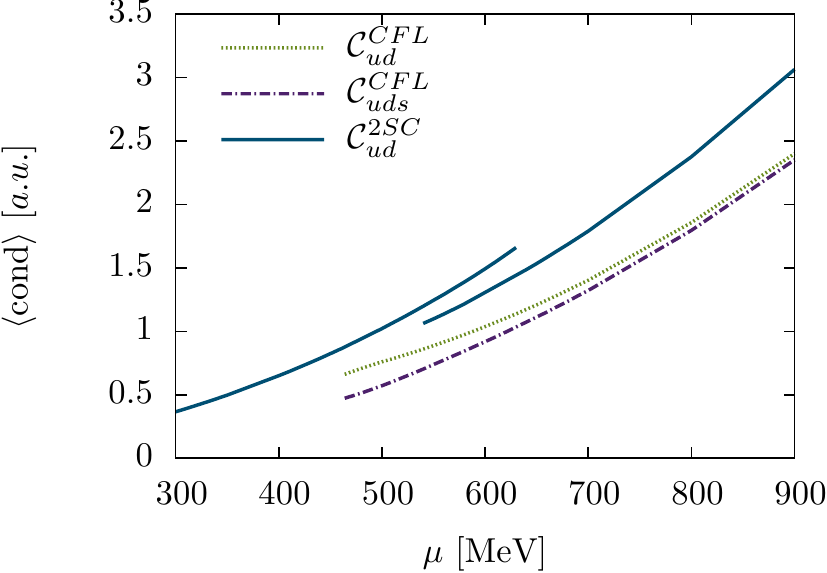}	
	\centering
		\includegraphics[]{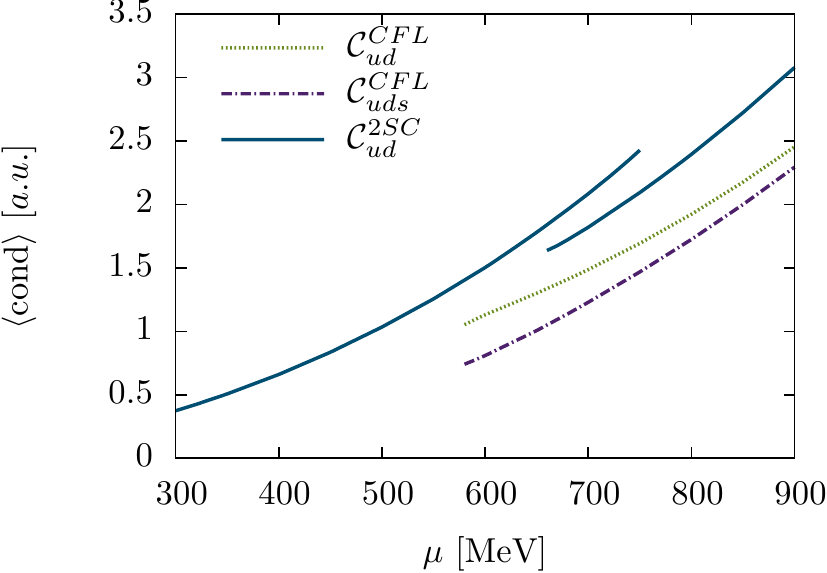}	
	\caption{2SC and CFL  solutions of the diquark condensates
	at $T=10$ MeV as functions of the chemical potential for $m_s=30$ MeV (top) and $m_s=54$ MeV (bottom).}
\label{fig:condmu}
\end{figure}
\begin{figure}
	\centering
		\includegraphics[]{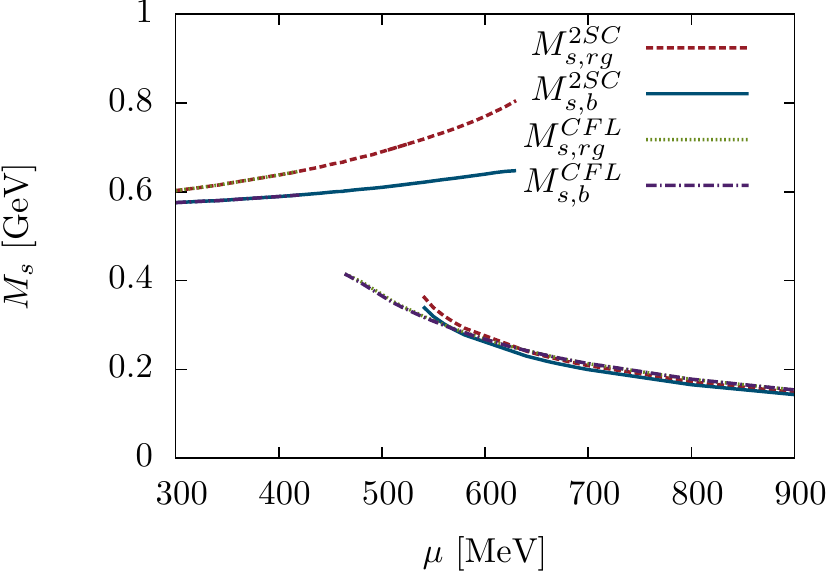}	
	\centering
		\includegraphics[]{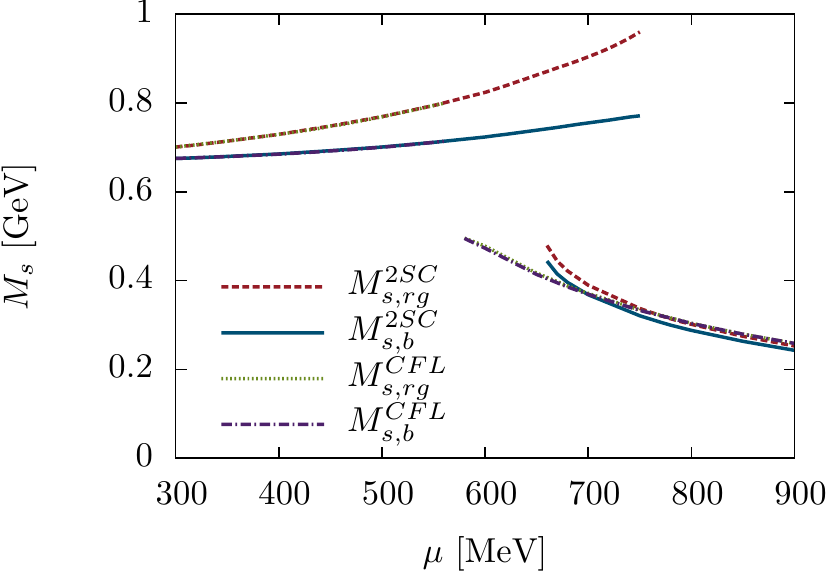}	
	\caption{Dependence of the dressed strange-quark masses at $T=10$ MeV on the chemical potential for $m_s=30$ MeV (top) and $m_s=54$ MeV (bottom).}
\label{fig:qmass}
\end{figure}

In \fig{fig:condmu} we show our solutions for the diquark condensates at $T=10$~MeV as functions
of the chemical potential. We have always indicated the corresponding pairing pattern, i.e., 2SC or CFL, 
to which these solutions correspond. In addition, there are sometimes different solutions with the same pairing pattern, which mainly differ by the dressed strange-quark mass, as shown in  \fig{fig:qmass}. Both figures should therefore be viewed together for a proper interpretation of the results.

The diquark condensates are qualitatively similar to the HTL-HDL case in our previous study. While the CFL condensates can only be formed with relatively light strange quarks, 2SC pairing is also possible for heavy strange quarks, as these quarks are not part of the condensates. However, due to the back-coupling, the strange quarks also have an influence on the 2SC condensates, as is evident from the discontinous behavior. This is the main difference to the HTL-HDL truncation.

The dressed strange-quark masses displayed in  \fig{fig:qmass} have been defined as the ratios
$M_s= (B_i^+(p)/C_i^+(p))|_{\vec p = \vec 0, n=0}$ of the Dirac components of the inverse propagator, \eq{eq:prop_dirac}. Therein we distinguish between red or green quarks, corresponding to the components
proportional to the color-flavor matrix $P_i= P_6$, and blue quarks, corresponding to $P_i= P_3$, cf.~\eq{eq:SPTM}.  Again we have also indicated the pairing pattern, to which the solutions belong. 

As already mentioned, the two branches of the 2SC solutions mainly correspond to different values of $M_s$.  
Although in the 2SC phase only up and down quarks are paired, the strange quarks have an impact on the diquark condensates through the polarization loop in the gluon propagator. In turn, the pairing leads to a visible difference between the masses of red/green and blue strange quarks in the 2SC phase, since the asymmetry between paired (red/green) and unpaired (blue) light quarks translates via the gluons also to the strange sector.
Both effects are not present in the simpler HDL-HTL truncation~\cite{Muller:2013pya}, where the gluon propagator is not affected by chiral and diquark condensates.

Above a threshold of around $\mu = 500$ MeV or $\mu = 600$ MeV for $m_s=30$ MeV or $m_s=54$, respectively, CFL pairing is possible. The CFL condensates show no qualitative difference to the HTL-HDL approximation and increase smoothly with rising chemical potential. In contrast to the 2SC solutions, the CFL quark masses of blue and red/green quarks do not differ markedly, as the pairing is almost symmetric.

\begin{figure}
	\centering
		\includegraphics[]{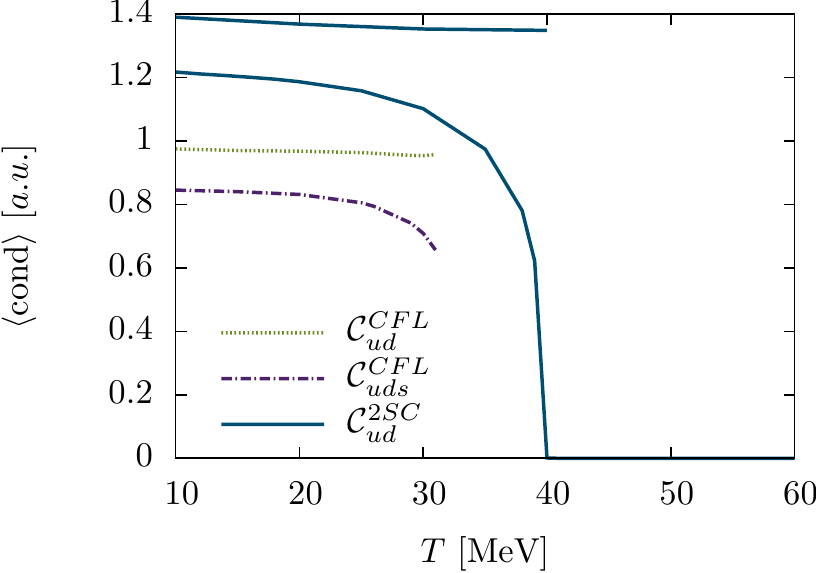}	
	\centering
		\includegraphics[]{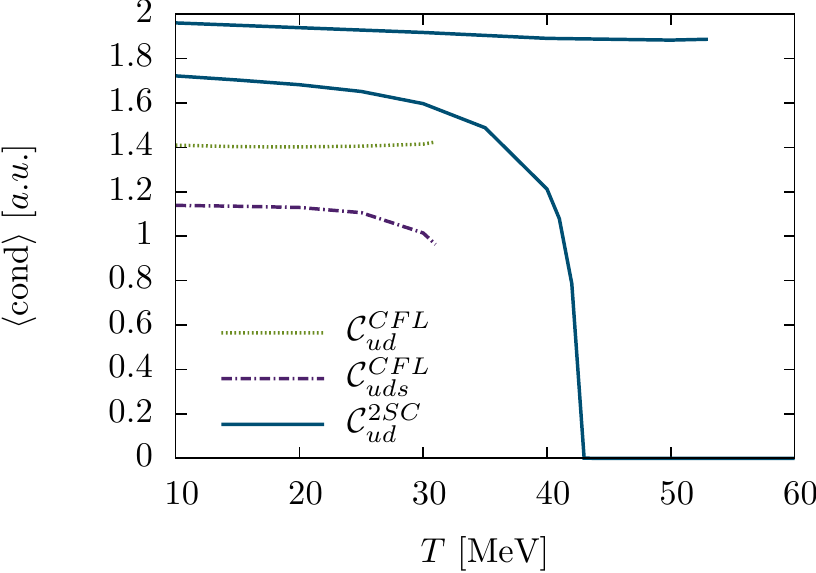}	
	\caption{Dependence of 2SC and CFL condensates on temperature for $m_s=30$ MeV at $\mu=580$ MeV (top) and for  $m_s=54$ MeV  at $\mu=680$ MeV (bottom).
	}
\label{fig:condT}
\end{figure}

The dependence of the condensates on temperature at a fixed chemical potential of $\mu=580$ MeV for $m_s=30$~MeV and of $\mu=680$ MeV for $m_s=54$~MeV is shown in \fig{fig:condT}. These chemical potentials are inside the region where a CFL solution and two branches of 2SC solutions exist. The upper and the lower branch correspond to the higher and the lower value of $M_s$, respectively. Similar to the HTL-HDL case, the 2SC solutions extend to higher temperatures than the CFL solution. The solution with the lower strange-quark mass goes smoothly down to  zero, suggesting a second-order phase transition to the normal-conducting phase at this point. An important quantitative difference is that the 2SC solutions exist up to critical temperatures of 
about $40-50$~MeV, while with HTL-HDL truncation we found critical temperatures of only about $20$~MeV~\cite{Muller:2013pya}.

\begin{figure}
	\centering
		\includegraphics[]{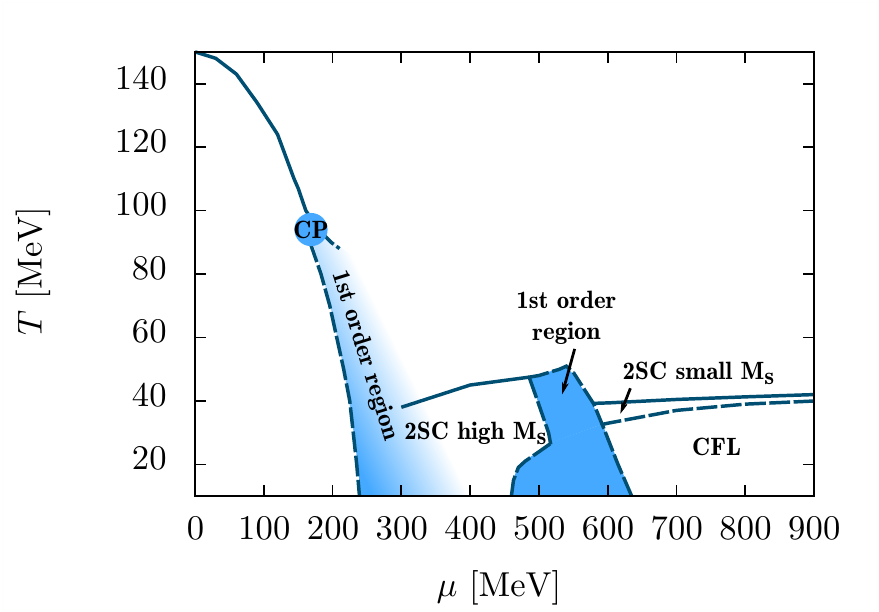}	
	\centering
		\includegraphics[]{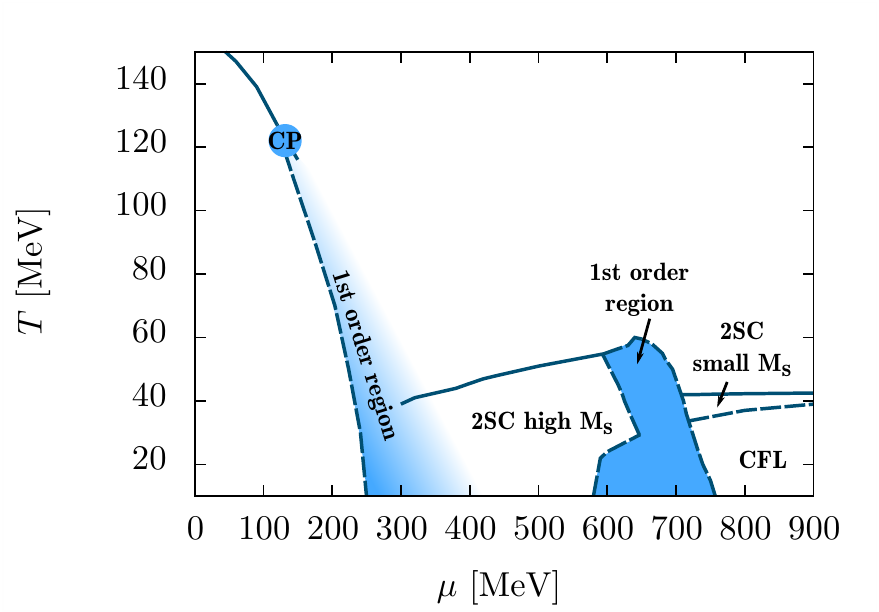}	
	\caption{Phase diagram for $m_s=30$ MeV (top) and $m_s=54$ MeV (bottom). First-order regions are indicated by shaded areas bounded by spinodal lines (dashed). Solid lines indicate second-order transitions, CP the tricritical point.}
\label{fig:pd}
\end{figure}

In order to determine the phase structure at given temperature and chemical potential, we should in principle compare the pressure of the different solutions,  which is given by the effective action from which the DSEs can be derived. In practice this usually turns out to be numerically very difficult, even when the analytical expression for the effective action is known. Unfortunately, in the present truncation scheme, we even do not have an analytic expression for the effective action.\footnote{Besides details of the regularization of the gluon polarization,
this is mainly related to our ansatz for the dressed quark-gluon vertex, which contains the anomalous components of the quark self-energy, see \eq{eq:bc_vertex}, but cannot be derived from a diagram. Strictly speaking, this means that the present truncation scheme is thermodynamically not fully consistent. However, in comparison with the HTL-HDL scheme, we believe that the merits of the improved truncation outweigh the possible errors related to this inconsistency. }
Hence, if we have more than one solution, we can in general not decide, which of them is favored. However, as motivated in~\cite{Muller:2013pya}, we expect that the stability of the numerical iteration is a measure of the thermodynamical stability of the phase. In particular, the iteration only converges to thermodynamically stable or metastable solutions, and the disappearance of the numerical solution at a certain temperature or chemical potential signals that this solution turns thermodynamically unstable at that point. This picture is corroborated by the fact that in \fig{fig:qmass} the onset of the 2SC solution with the lower $M_s$ is at a higher chemical potential
than the onset of the CFL solution: The energy gain due to CFL condensation can exceed the energy gain due to the dynamical mass generation and therefore a solution with lower mass gets stabilized earlier.  

According to the above assumption, second-order phase transitions can be located precisely,
while in the case of first-order phase transitions we can only determine the spinodal regions,
i.e., the regions where two or more at least metastable solutions coexist. 
First-order transitions between these solutions are then restricted to the spinodal region.
In this context, an additional complication  arises from the impact of the strange-quark phase transition on the non-strange sector. As a consequence, sometimes more than two
solutions coexist, so that it is not even always clear between which phases a phase transition takes
place in the spinodal region. 
Therefore the phase diagrams, which are displayed in \fig{fig:pd}, have a more complicated structure 
with much larger spinodal regions than in the HTL-HDL case shown in Ref.~\cite{Muller:2013pya}.
Especially the spinodal region of the chiral first-order transition (below the tricritical point indicated
by ``CP'') becomes remarkably large,
and we were not even able to find the exact position of the upper spinodal anymore. 
Therefore the region is indicated only approximately. 

The chiral phase transition seems to be quite robust under the variation of the bare strange-quark mass, although the critical point is shifted to higher temperatures and lower chemical potentials when increasing $m_s(\nu)$. We find the critical point around $(T,\mu)=(100,160)$~MeV for $m_s=30$ MeV and $(T,\mu)=(120,120)$~MeV for $m_s=54$ MeV. A similar study in~\cite{Fischer:2012vc} finds a critical endpoint at $T=100$ MeV and $\mu=190$ MeV and sees the same qualitative change in comparison with the HTL-HDL approximation. However, it should be noted that the regularization of the quark loop and the vertex truncation were done in a slightly different way in that work.

As in Ref.~\cite{Muller:2013pya}, our main focus lies on the color-superconducting phases at higher chemical potentials. We find a CFL-like phase at high $\mu$ and a 2SC phase at intermediate $\mu$ as well as in a small band at intermediate temperature, separating the CFL phase and the normal-conducting phase. Qualitatively, this phase structure is similar to the HTL-HDL case. On the other hand, there are some important differences.
The coupling of the strange quarks to the light sector leads to a splitting of the 2SC phase into two phases which are not continuously connected: a phase where $M_s$ is relatively large and a phase where it is considerably smaller. Therefore, we find a large spinodal region where four phases - the CFL phase, the two 2SC phases and the normal-conducting phase - meet, and all phase transitions between these phases are of first order.
In the lower-temperature part of this region, the normal-conducting phase becomes unstable with respect to the color-superconducting phases (not explicitly indicated in the phase diagram). Although we cannot make a definite statement, we expect most of this lower-temperature part to be in the CFL phase, as the 2SC solutions are numerically much less stable.\footnote{This means that for most starting values the iterative solution of the DSE converges to the CFL solution, whereas the 2SC solution is only found if the starting values are already very close to this solution.} This would mean that the strange-quark transition at low temperature is shifted to lower chemical potentials than it would occur within the 2SC or the normal conducting phase. Again, this can be explained by the additional condensation energy due to the CFL pairing.  

Except for the spinodal region, the transition between 2SC and normal-conducting phase is of second-order and takes place between $T=40$ and $60$ MeV. In general, the critical temperature rises with increasing chemical potential, but drops suddenly at the strange-quark phase transition.

Although the bare strange-quark mass has some influence on the critical point of the chiral phase transition, the color-superconducting section and especially the 2SC phase are quite independent of $m_s$. The main difference is that the onset of CFL condensation, together with the  phase transition between the two 2SC phases, is shifted to larger chemical potentials for the larger $m_s$. The critical temperature to the normal-conducting phase 
only shows a weak dependence on the $m_s$. 
 

\section{Debye and Meissner masses}\label{sec.debye}

\begin{figure}
	\centering
		\includegraphics[]{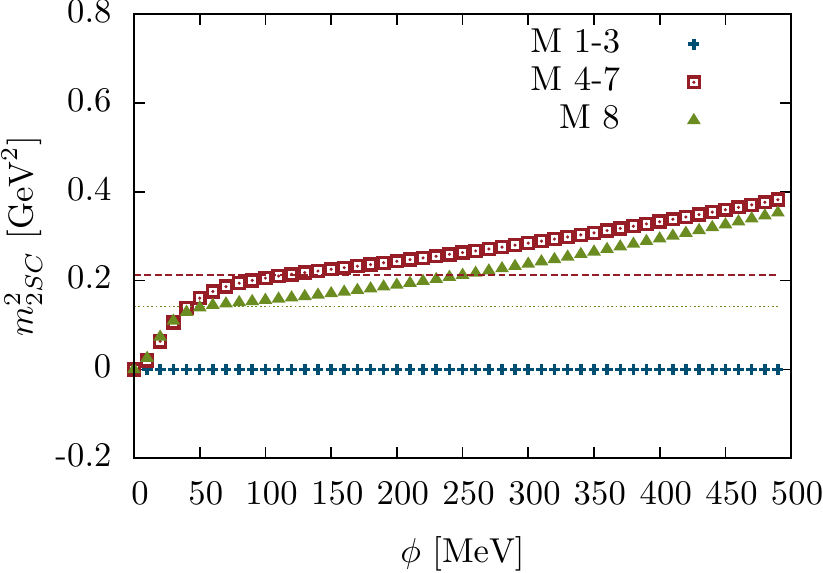}	
	\centering
		\includegraphics[]{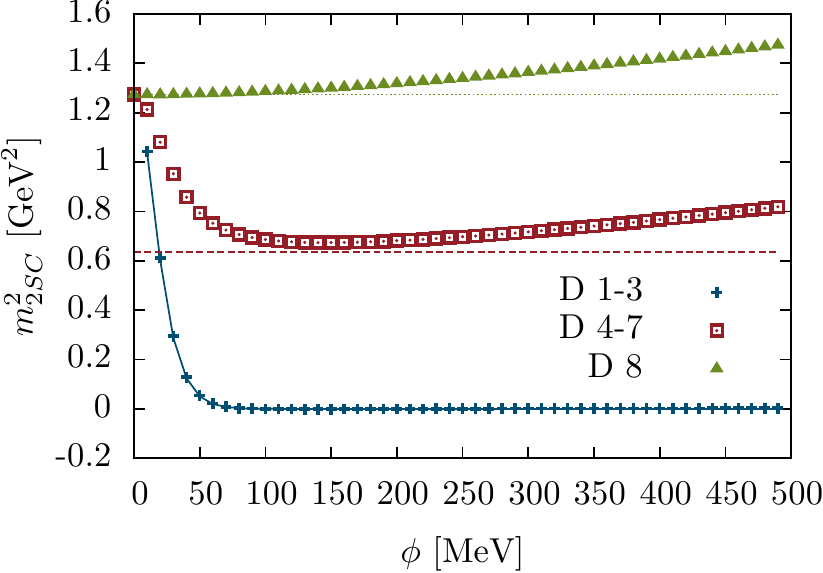}	
	\caption{Meissner (top) and Debye (bottom) masses (dots) in comparison with weak-coupling 
	results~\cite{Rischke:2000qz} (lines) for the 2SC phase for propagators \eq{eq:propsimp}.}
\label{fig:glumass_wc_2sc}
\end{figure}
\begin{figure}
	\centering
		\includegraphics[]{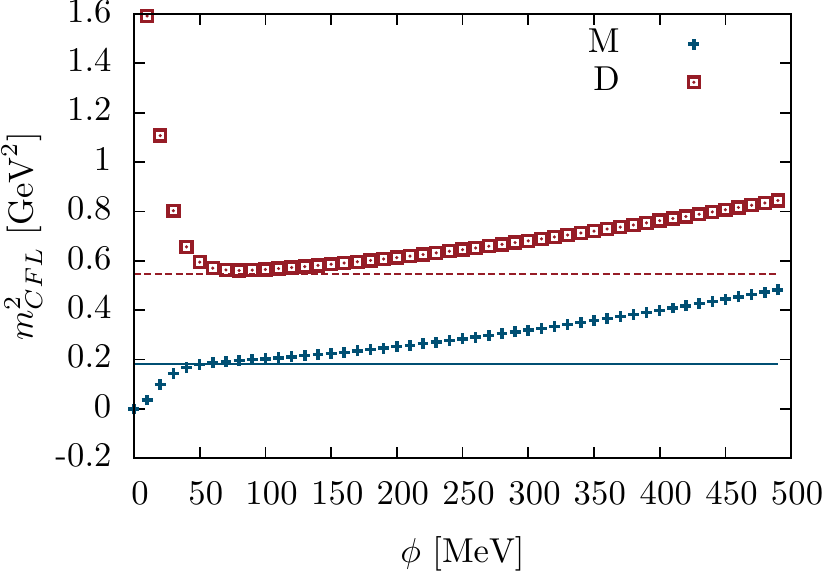}	
	\caption{Meissner and Debye masses (dots) in comparison with weak-coupling results~\cite{Rischke:2000ra} (lines) for the CFL phase for propagators \eq{eq:propsimp}.}
\label{fig:glumass_wc_cfl}
\end{figure}

Due to dressing, the gluons acquire effective masses. These are defined by the full gluonic polarization tensors at zero momentum. Here we focus on the quark contribution to the screening masses defined by\footnote{As we do not solve the gluon DSE and do not know the value of the gluon renormalization constant $Z_3$, these masses are strictly speaking not the Debye and Meissner masses but renormalization point dependent objects, which depend on $Z_3$. They are a qualitative measure for the Debye and Meissner masses but cannot be compared quantitatively with the HTL-HDL results.}:
\bseq
m_{D,ab}^2 \= \lim_{\vec p \rightarrow 0}\Pi^{ab}_{TL}(\omega_m=0,\vec p) \\
m_{M,ab}^2 \= \lim_{\vec p \rightarrow 0}\Pi^{ab}_{TT}(\omega_m=0,\vec p)
\eseq
These masses are called Debye and Meissner masses and account for the electric and magnetic screening of the gluons. For non-superconducting phases the Meissner masses are equal to $0$ due to the unbroken gauge symmetry. For color superconductivity, the $SU_c(3)$ color symmetry is spontaneously broken. The resulting Goldstone bosons are ``eaten'' by the gluons giving rise to magnetic gluon masses via the Anderson-Higgs mechanism \cite{Anderson:1963pc,Higgs:1964pj}. For the 2SC phase the $SU_c(3)$ is broken down to $SU_c(2)$ giving rise to 5 massive gluons while the remaining 3 gluons stay massless. For the CFL phase the whole $SU_c(3)$ group is broken and hence all 8 gluons acquire Meissner masses.

Before presenting the results for our coupled DSE system, we check the validity of our truncation by comparing
the Debye and Meissner masses with weak-coupling results given in \cite{Rischke:2000qz,Rischke:2000ra}. 
To this end, we use simple propagators with selfenergies 
\beq\label{eq:propsimp}
\Phi^+ = \phi_{i}\gamma_5 M_{i},
\eeq
as in \eq{eq:cssimple}, but with constant gaps $\phi_i$.  For the color-flavor structure, determined by the matrices $M_i$, we consider a 2SC phase with $N_f=2$, i.e., $M_i = M_{2SC}$, see \eq{eq:phi2sc}, as well as a CFL phase with $N_f=3$ massless quarks, i.e., $M_i = M_{oct/sing}$, see \eq{eq:phicfl}.

The results for the 2SC phase at $T=10$ MeV and $\mu=1000$ MeV are shown in \fig{fig:glumass_wc_2sc},
where the squared Debye and Meissner masses for the different gluons are displayed as functions of the gap parameter $\phi = \phi_{2SC}$ and compared with the weak-coupling results of \cite{Rischke:2000qz}.
In addition to the simple propagator parametrization, the weak-coupling limit assumes $T\ll \phi \ll \mu$. This is fulfilled best for $\phi$ around $100$ MeV where we find good agreement between the weak-coupling results and our results. For larger and smaller values there are some deviations. For the Debye mass of gluons 1-3 a temperature dependent weak-coupling result is provided in Ref.~\cite{Rischke:2000qz}, which coincides exactly with our calculation. The Meissner mass of the gluons 1-3 vanishes exactly, as they correspond to the $SU_c(2)$ subgroup which stays unbroken in the 2SC phase. In addition, the Debye mass of gluon 1-3 also tends towards $0$ for low temperature. These gluons can only couple to red or green quarks, which are all bound in Cooper pairs for 2 flavors. Therefore, the quark loop gives no contribution and no Debye mass is generated. All other gluons acquire both, Debye and Meissner masses.

The corresponding results for the CFL phase are shown in \fig{fig:glumass_wc_cfl} as functions of
 $\phi=\phi_{oct}=\frac{1}{2}\phi_{sing}$. As the breaking pattern is symmetric in color and flavor, all gluons acquire the same Debye and Meissner masses. The weak-coupling results \cite{Rischke:2000ra} are shown again for comparison and agree reasonably well in the region where $T\ll \phi \ll \mu$. It is worth to mention that the off-diagonal vertex contributions \eq{eq:bc_vertex} are essential to reproduce the weak-coupling results.

\begin{figure*}
	\centering
		\includegraphics[]{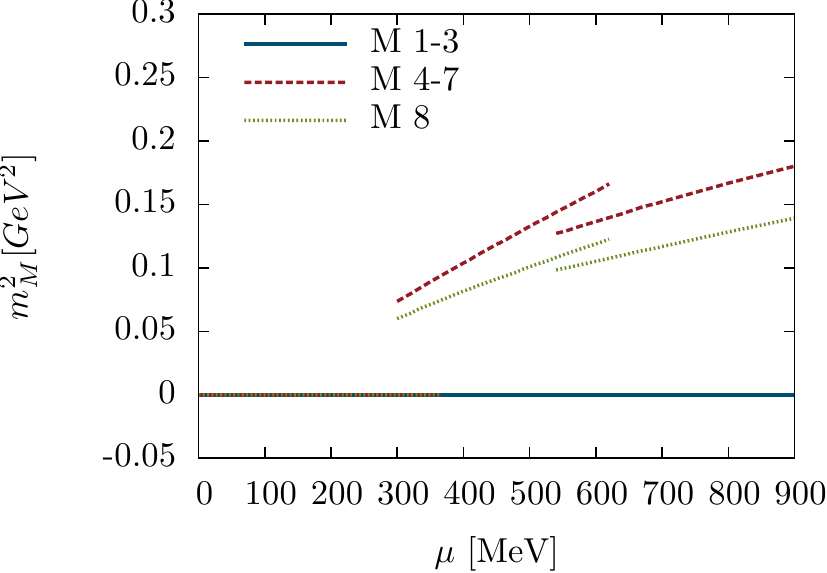}	
	\centering
		\includegraphics[]{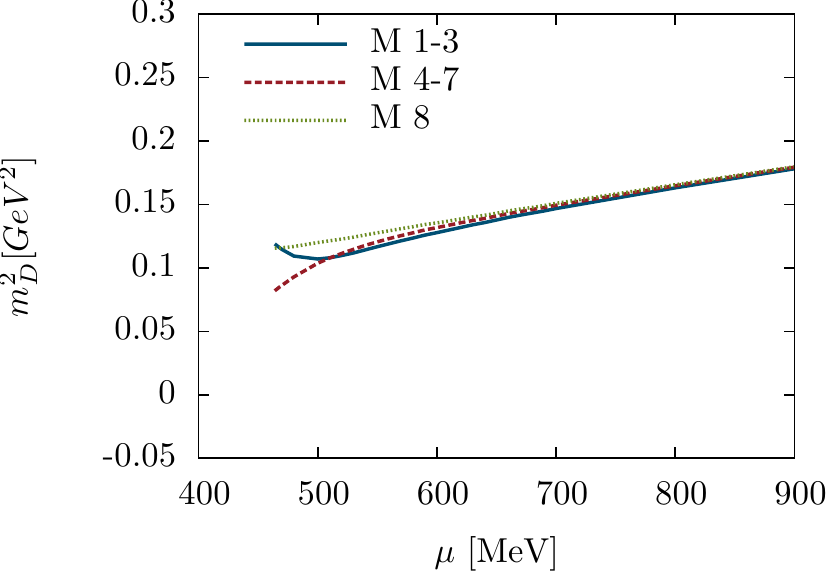}
	\centering
		\includegraphics[]{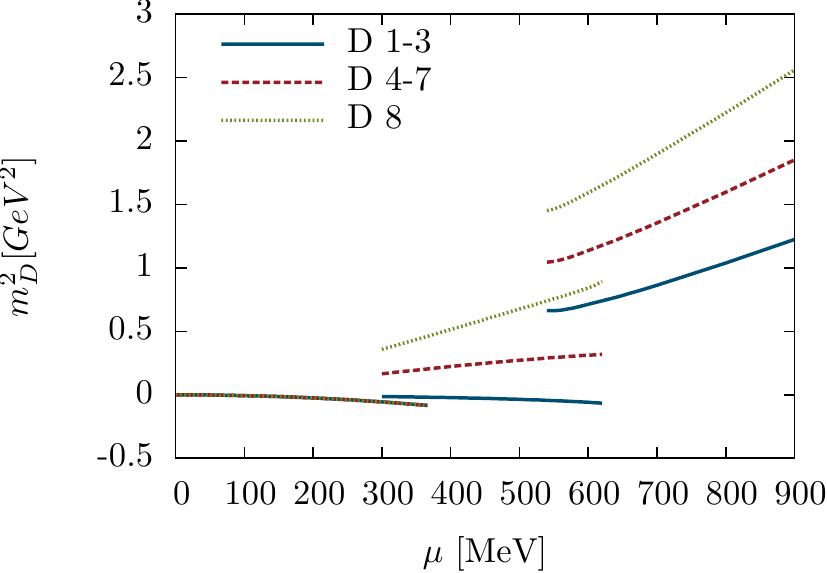}	
	\centering
		\includegraphics[]{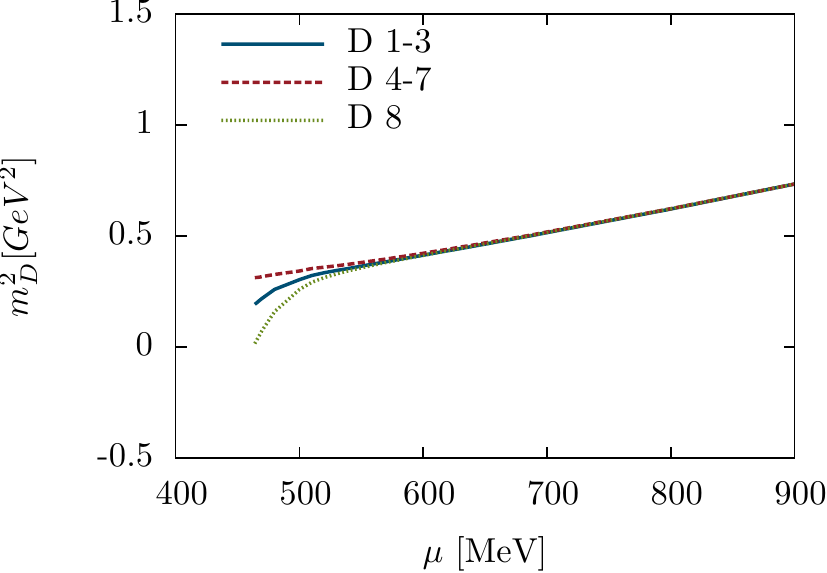}		
	\caption{Meissner (top) and Debye (bottom) masses squared for the 2SC phase (left) and the CFL phase (right) for $m_s=30$ MeV.}
\label{fig:glumass_1}
\end{figure*}

\begin{figure*}
	\centering
		\includegraphics[]{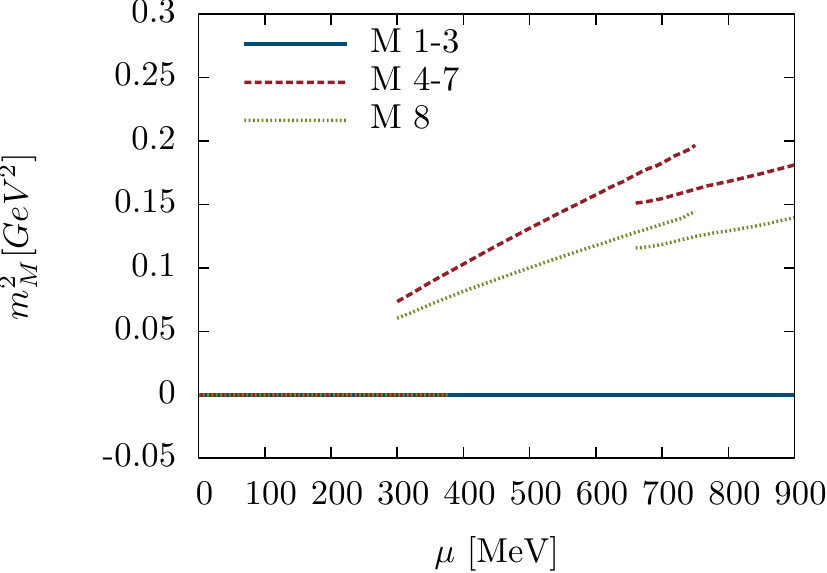}	
	\centering
		\includegraphics[]{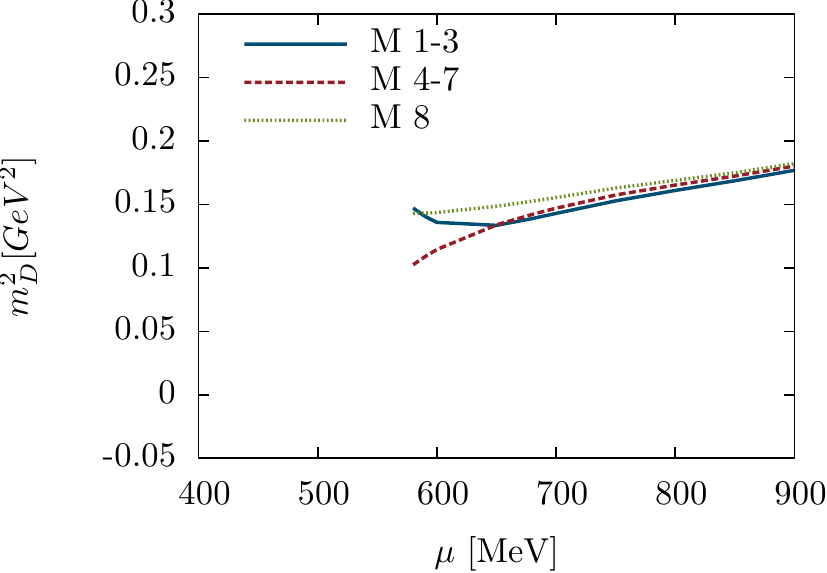}	
		\centering
		\includegraphics[]{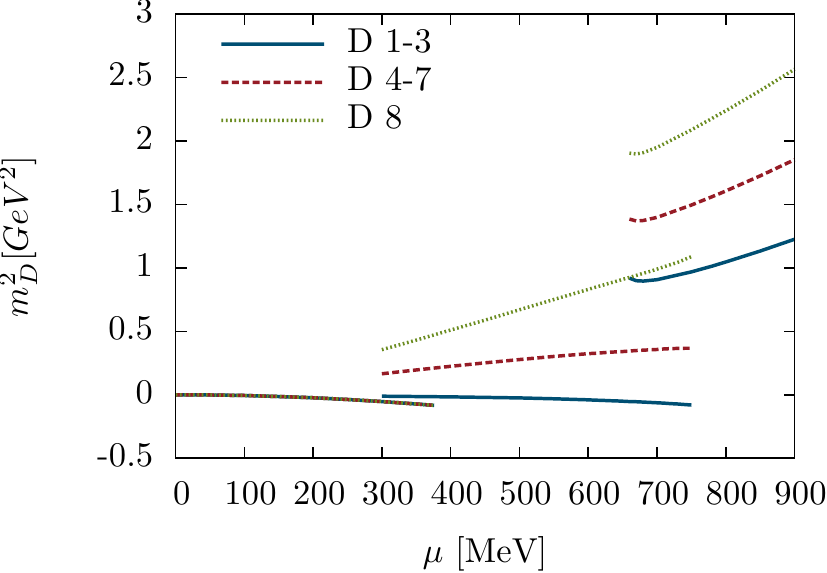}	
	\centering
		\includegraphics[]{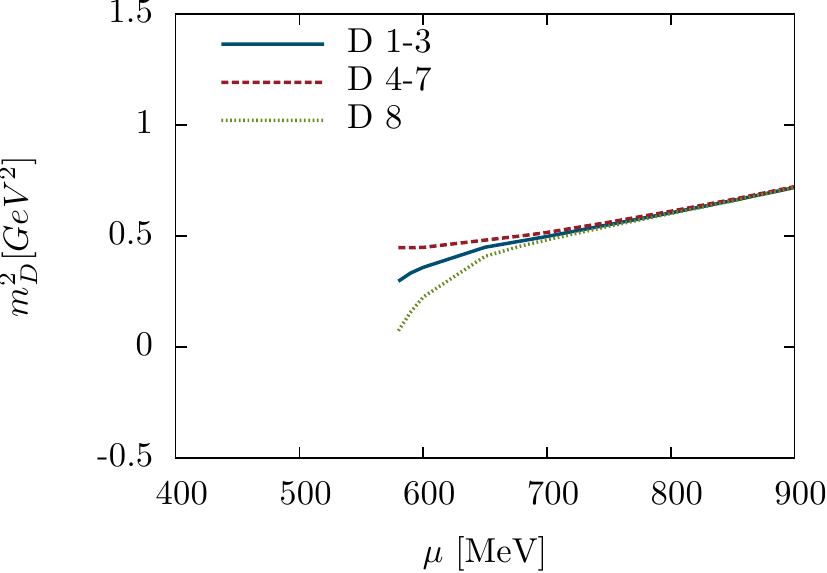}		
	\caption{Meissner (top) and Debye (bottom) masses for the 2SC phase (left) and the CFL phase (right) for $m_s=54$ MeV}
\label{fig:glumass_2}
\end{figure*}

Finally we also show the Debye and Meissner masses for the full DSE calculations for $m_s=30$ MeV in \fig{fig:glumass_1} and $m_s=54$ MeV in \fig{fig:glumass_2}. At low chemical potentials all masses are small as the Debye masses are suppressed by the heavy quark masses and Meissner masses are zero in non-color-superconducting phases. In this region, we actually find that the squared Debye mass slighly decreases and becomes negative. Small negative squared Debye masses are not forbidden in principle, as there is always a positive Yang-Mills contribution to the Debye mass, so that the sum of Yang-Mills and quark contribution can still be positive. On the other hand, the observed behavior also violates the so-called Silver-Blaze property, which requires observables and also the Debye mass at $T=0$ to be independent of the chemical potential, if the latter is smaller than the mass gap of the theory. Although the calculations have been performed at $T=10$~MeV, 
the mass change is mostly an artifact of the vertex approximation. This will be discussed in detail in the 
next section.

At $\mu=300-400$ MeV, there is a first-order phase transition to the 2SC phase. Similar to the weak-coupling results, gluons 1-3 have no or only small Meissner and Debye masses, and the relative ordering of the gluon masses is the same as in the weak-coupling approximation. At higher chemical potential also the strange quarks undergo a phase transition and become light. Therefore, the strange quark loop gives larger contributions and increases the Debye masses of all gluons equally while it does not contribute to the Meissner masses. 

CFL pairing is possible for $\mu$ larger than $450$ or $550$ MeV for $m_s=30$ or $m_s=54$ MeV respectively. In this phase, the Debye and Meissner masses of all 8 gluons become similar due to the symmetric pairing pattern, like in the weak-coupling limit. The deviations originate from the finite strange-quark mass and diminish with increasing $\mu$, as the mass becomes negligible with respect to the chemical potential. Both values of $m_s$ give very similar results, the main difference being the position of the strange quark phase transition.

\section{Silver-Blaze property}
\label{silverblaze}

The Silver-Blaze property (SBP)~\cite{Cohen:2003kd,Cohen:2004qp} states that in a relativistic system at zero temperature, the partition function and observables do not depend on chemical potential if the latter stays below the mass gap $\Delta$ of the system. Although the Lagrangian of the theory and various related quantities  (like propagators) depend on $\mu$, observables, such as the pole masses of physical particles, must stay constant, i.e., the internal $\mu$-dependencies must cancel each other. If the chemical potential exceeds the mass gap, states can be excited and observables may change. The SBP only holds at zero temperature, as thermal excitations also change observables.

In strong-interaction matter at $T=0$ and nonzero baryon chemical potential $\mu_B = 3\mu$, the physical threshold is given by $\mu_B = m_N-E_b$, where $m_N$ is the nucleon mass and $E_b = 16$~MeV is the binding energy in nuclear matter. Since nucleons and nuclear binding are not explicitly contained in the present truncation scheme, the threshold should be the transition point to deconfined quark matter, i.e.,  the phase transition to the 2SC phase. 

In contrast to this expectation, the Debye masses shown in Figs.~\ref{fig:glumass_1} and
\ref{fig:glumass_2} are $\mu$ dependent at arbitrarily small chemical potentials. As mentioned earlier, the calculations have been performed at a finite temperature of $T=10$~MeV, but this is too small to explain the observed variations of the Debye masses. On the other hand, the SBP can easily be violated by truncations. 
In the following, we discuss this in more detail. 

We consider a system with a physical mass gap $\Delta$, meaning that the real-time quark propagator 
as well as the gluon propagator and the quark-gluon vertex in vacuum do 
not have non-analytic structures for energies lower than $\Delta$. Turning to the Matsubara formalism, the Euclidean momentum component $p_4 = \omega_n + i\mu$ becomes continuous at $T=0$.
Writing $p_4 = \tilde p_4 +i\mu$ and suppressing the dependence on $\vec p$, the quark self-energy is then schematically given by
\beq
\Sigma(\tilde p_4+i\mu) \sim \int_{\tilde q_4} S(\tilde q_4+i\mu) K(\tilde p_4+i\mu,\tilde q_4+i\mu)\,,
\eeq
with an integration kernel $K(p_4,q_4)$. By the assumption specified above, the integrand is analytic in the entire region between $\mathrm{Im}\,\tilde q_4 = 0$ and $\mathrm{Im}\,\tilde q_4 = \Delta$. Hence, if $\mu < \Delta$,  the substitution $\tilde q_4 +i\mu \rightarrow \tilde q_4$ and shifting the path of integration back to the real axis does not alter the value of the integral, i.e., we obtain
\beq
\label{eq:sb_selfen}
\Sigma(\tilde p_4+i\mu)  \sim \int_{\tilde q_4} S(\tilde q_4) K(\tilde p_4+i\mu,\tilde q_4)\,,
\eeq
which depends on the chemical potential only through the external energy variable. The selfenergy is therefore given by the same function as in vacuum, $\Sigma(\tilde p_4+i\mu) \equiv \Sigma_{vac}(\tilde p_4 + i\mu)$, 
and the chemical potential only determines the complex argument at which this function is to be evaluated. 
For chemical potentials higher than $\Delta$, on the other hand, the singularity caused by this mass gap prevents the shift of the integration path, and therefore the selfenergy is not simply an analytical continuation of the vacuum function.

Obviously, the same holds for the dressed quark propagator, which is related to $\Sigma$ via the DSE. For $\mu<\Delta$ an analogous shift then yields
\beq
\int_{\tilde q_4} S(\tilde q_4+i\mu) = \int_{\tilde q_4} S(\tilde q_4), 
\eeq
i.e., this integral, which arises when calculating the quark condensate, is independent of $\mu$. 
Hence, while the quark propagator itself has a $\mu$-dependence (via its energy argument, $S(q_4) \equiv S(\tilde q_4 + i\mu)$), the condensate is constant, reflecting the Silver Blaze property. 

Similarly, the gluon polarization integral, schematically given by
\beq
\Pi(k_4)\sim \int_{\tilde q_4} S(\tilde q_4+i\mu)S(\tilde p_4+i\mu) \tilde K(\tilde p_4+i\mu,\tilde q_4+i\mu),
\eeq
with $\tilde p_4=k_4 + \tilde q_4$, is independent of $\mu$ for $\mu<\Delta$, and therefore also Debye and Meissner masses are constant.

A technical requirement for these properties to hold is that the integrands, i.e., the quark propagators and the kernels $K$ or $\tilde K$, consistently depend on the quark momenta shifted by $i\mu$, so that the $\mu$-dependence can be removed by performing a shift of the integration variable. Indeed, this is always fulfilled if the kernel is microscopically calculated from Feynman diagrams.

In our calculations, however, we make the model ansatz $\Gamma(p,q) = \Gamma(\kappa^2)$,
 \eq{eq:vertex1}, so that it depends on the choice of $\kappa$ whether the SBP is preserved. In the quark DSE, we use $\kappa^2 = (p-q)^2$, which is consistent with the SBP. For the gluon-polarization loops, on the other hand, we take $\kappa^2 = \tilde p^2 + \tilde q^2 = \tilde p_4^2 + \vec p\,^2 + \tilde q_4^2 + \vec q\,^2$,
which violates the SBP, since $\tilde p_4$ and $\tilde q_4$ are not shifted by $i\mu$. As discussed in Sec.~\ref{sec.trunc.qgv}, we made this choice, because the formally more consistent ansatz $\kappa^2 = p^2 + q^2$
would lead to unphysical singularities in the vertex function, which we consider to be a more severe problem
than the vioalation of the SBP.

\begin{figure}
	\centering
		\includegraphics[]{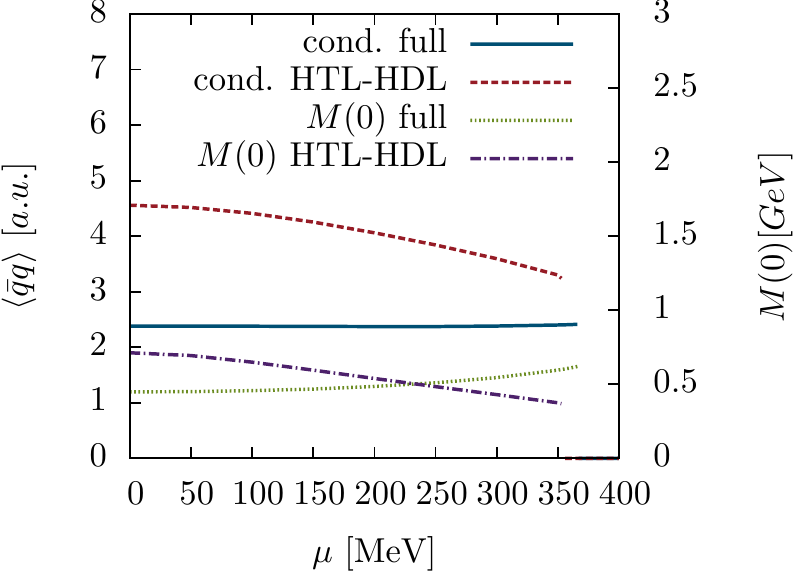}	
	\caption{Light quark condensate and light-quark mass $M(0)=B_1(0)/C_1(0)$ for HTL-HDL truncation and the full back-coupling at $T=10$~MeV.}
	\label{fig:ch_cond}
\end{figure}

In fact, the violation is negligible on the quark level, as can be seen in \fig{fig:ch_cond}. The quark condensate calculated within the present truncation scheme (blue solid line) stays almost constant as a function of $\mu$, showing only a tiny increase near the phase transition. For comparison we also show the condensate in the HTL-HDL truncation scheme (red dashed line). In HTL-HDL approximation, the gluon propagator is is dressed by massless quarks, giving the gluons an effective mass  $m_g^2\sim \mu^2$. This causes a strong violation of the SBP, which is clearly visible in the figure. Hence, taking into account the quarks in the gluon polarization self-consistently leads to a large improvement, while the violation of the SBP by the vertex ansatz is negligible.

In the figure we also show the light quark ``mass'' $M(0) \equiv (B_1(p)/C_1(p))|_{\vec p = \vec 0, p_4=\pi T + i\mu}$ for both truncation schemes. The $\mu$ dependence which is visible in both cases does not allow to make a statement about the SBP because $M(0)$ is not an observable, and the quark selfenergy depends on $\mu$ through its momentum argument, cf.~\eq{eq:sb_selfen}.

For the Debye masses of the improved truncation, \fig{fig:glumass_1} and \fig{fig:glumass_2}, the violation of the SBP is a bit larger than for the quark condensate, but still on a tolerable level if we compare its size with the physical effects in the 2SC or CFL phases. Altogether we can safely state that the improved truncation respects the SBP approximately, with the violations being negligibly small in the quark sector and not too big in the gluonic sector.

\section{Conclusions and Outlook}
\label{sec:conclusions}
Within the Dyson-Schwinger approach we have studied 2SC and CFL-like phases in QCD with $2+1$ flavors with chiral light quarks and strange quarks of finite bare mass $m_s$. To explore the sensitivy of the results,  
two values, $m_s=30$~MeV and $m_s=54$~MeV at a renormalization scale of $\nu=100$~GeV,
were chosen. Going beyond our previous work~\cite{Muller:2013pya}, which employed the HTL-HDL approximation, the quark effects on the gluon propagator are now taken into account selfconsistently through fully dressed Nambu-Gorkov quark propagators. The adopted truncation and regularization schemes were inspired by a Brown-Pennington projection and Slavnov-Taylor identities. Guided by the most up-to-date lattice results for the equations of state at vanishing chemical potential~\cite{Borsanyi:2010bp,Bazavov:2014pvz} we have fitted the vertex parameters to a chiral critical temperature of $150$~MeV for both values of $m_s$. As an improvement to~\cite{Muller:2013pya} 
this also yields a good description of the pion decay constant in vacuum, and
we now obtain more realistic vacuum values of the dressed light-quark masses with little sensitivity to the choice of $m_s$. 

For the color-superconducting phases the critical temperatures go up to $40-60$ MeV. 
In contrast to the HTL-HDL approximation, the light-quark sector is now coupled to the strange sector through the 
quark loops in the gluon polarization function.
As a consequence we find two distinct 2SC phases, 
separated by a first-order phase transition with a discontinuous change in the dressed strange-quark mass.
The main effect of increasing the bare strange-quark mass $m_s$ is shifting the 2SC-CFL boundary to higher $\mu$
and the location of the chiral critical point to higher $T$ and lower $\mu$.
The critical temperatures from the color superconducting to the normal conducting regime, on the other hand,
show only a weak dependence on $m_s$.

We have also calculated the gluonic Debye and Meissner masses 
in the different color channels and phases.  
As a test for our truncation scheme, we first calculated the masses in the weak-coupling limit 
and found agreement with the corresponding predictions in the literature~\cite{Rischke:2000qz,Rischke:2000ra}.
To achieve this a proper construction of the quark-gluon vertex, containing anomalous contributions, was crucial.
The results of the full calculations in the strong-coupling regime are qualitatively similar.
Especially the gluonic masses have the same hierarchy as in the weak-coupling limit. 

Although the selfconsistent treatment of the quark propagators in the gluon polarization is a significant improvement to the HTL-HDL approximation, there are still some features missing in the truncation. The gluon propagator can be further improved by including quark effects to the Yang-Mills sector as these may have an influence on the Yang-Mills screening mass. Even more important would be an improvement of the vertex truncation. 
Although the vertex we use is partially constrained by Slavnov-Taylor identities and the correct UV behavior,
it contains nevertheless a consideral amount of modeling.
This does not only reduce the predictive power of the approach but also leads to artifacts, like the violation
of the Silver-Blaze property.
It would thus be desireable to calculate the vertex function explicitly from the corresponding DSE. 
This has already been studied in vacuum~\cite{Alkofer:2008tt, Fischer:2009jm, Hopfer:2013np},
but unfortunately becomes much more difficult in the medium.  
Finally, a truncation that provides an analytic expression for the effective action should be considered in the future, as it allows to calculate pressure differences between the phases and therefore the location of the first-order transitions in the phase diagram.

\section{Acknowledgements} 
We would like to thank Christian Fischer, Jan L\"ucker and Dirk Rischke for interesting discussions and helpful comments. D.M. was supported by BMBF under contract 06DA9047I and by the Helmholtz Graduate School for Hadron and Ion Research. M.B. and J.W. acknowledge partial support by the Helmholtz International
Center for FAIR and by the Helmholtz Institute EMMI.

\appendix

\section{Parametrization of propagators and condensates}\label{app:par}

The matrices $P_i$ and $M_i$, $i= 1, \dots, 7$,  which para\-met\-rize the color-flavor structure of 
the quark propagator (cf. \eq{eq:SPTM}) and self-energies, are given in the color-flavor basis
$\lbrace(r,u),(g,d), (b,s),(r,d),(g,u), (r,s),(b,u),(g,s),$ $(b,d)\rbrace$ by
\beq\label{eq:pmcfl}
P_i = \left(\begin{smallmatrix}
\delta_{i,1}+\delta_{i,2} & \delta_{i,2} & \delta_{i,4}&&&&&&\\
\delta_{i,2} & \delta_{i,1}+\delta_{i,2} & \delta_{i,4}&&&&&&\\
\delta_{i,5} & \delta_{i,5} & \delta_{i,3}&&&&&&\\
&&&\delta_{i,1}&&&&\\
&&&&\delta_{i,1}&&&&\\
&&&&&\delta_{i,6}&&&\\
&&&&&&\delta_{i,7}&&\\
&&&&&&&\delta_{i,6}&\\
&&&&&&&&\delta_{i,7}\\
\end{smallmatrix}\right),
\eeq

\beq\label{eq:mmcfl}
M_i = \left(\begin{smallmatrix}
\delta_{i,1}+\delta_{i,2} & \delta_{i,2} & \delta_{i,4}&&&&&\\
\delta_{i,2} & \delta_{i,1}+\delta_{i,2} & \delta_{i,4}&&&&&\\
\delta_{i,5} & \delta_{i,5} & \delta_{i,3}&&&&&\\
&&&&\delta_{i,1}&&&\\
&&&\delta_{i,1}&&&&\\
&&&&&\delta_{i,7}&&\\
&&&&\delta_{i,6}&&&\\
&&&&&&&\delta_{i,7}\\
&&&&&&\delta_{i,6}&\\
\end{smallmatrix}\right).
\eeq

This basis allows for a consistent parametrization of CFL-like pairing for 2+1 flavors, where the corresponding amplitude functions are all different in general. On the other hand there are simplifying cases, where a smaller number of terms is sufficient. In the 2SC phase the anomalous self-energy $\Phi^+$ is proportional to a single color-flavor matrix given by
\beq
M_{2SC} = M_1-M_2
\label{eq:phi2sc}
\eeq
while the CFL phase in the limit of equal quark masses can be parametrized in terms of two
matrices 
\begin{align}
M_{oct} &= M_1+M_6+M_7-\frac{1}{3}\left(M_2-2M_3+M_4+M_5\right), \nonumber \\
M_{sing} &= -\frac{2}{3}\left(M_2+M_3+M_4+M_5\right)\,.
\label{eq:phicfl}
\end{align}

\section{Numerical details}\label{app:num_calc}

In Section \ref{sec.trunc.glu}, we have presented our truncation scheme of the gluon DSE. A direct calculation of the regularized quark polarization loop \eq{eq:bp_reg} is, although in principle finite, still numerically unstable, as it includes a complicated cancellation of the divergencies which cannot be achieved numerically.
We therefore subtract additional regulators $\Pi_{i,sub}= \tilde\Pi_i(0) - \tilde\Pi_L(0)$ where $\tilde \Pi_i$ corresponds to the vacuum expression for the polarization function $\Pi_i$, but evaluated with quark propagators $\tilde S(p_4,\vec p)$, where the real parts of the in-medium results for the $A$, $B$ and $C$ functions are used and interpolated to continuous $p_4$. The regularized polarization functions are therefore
\beq
\Pi_{i,reg}(k)=\Pi_i(k) - \Pi_L(0) - \lb \Pi_i(0) - \Pi_L(0)\rb_{T=0,\mu=0}, 
\eeq
for both components $i=TT/TL$.
If all terms have the same integration nodes for large integration momenta, this expression converges numerically, independent of the coarseness of the nodes, as there is always a pair of terms $\Pi-\Pi_{T=0,\mu=0}$ which cancel each other's divergencies. If $A_i(q)=C_i(q)$, as for example in vacuum, the subtracted terms $\Pi_{i,sub}$ can be shown to be zero analytically and serve solely for numerical stabilization. As $A_i(q)$ and $C_i(q)$ differ in the medium, the $\Pi_{i,sub}$ terms are not exactly zero and lead to a small truncation error. Although the chosen regulators seem to be a bit arbitrary, we have tested that the result only slightly changes for variations of the regularization, like setting also $B_i(q)=0$ or using $\frac{1}{2}(A_i(q)+C_i(q))$ instead of $A_i(q)$ and $C_i(q)$ in $\tilde S(p_4,\vec p)$. This means that the different IR parts of the dressing functions only have a negligible contribution while the similar UV parts are stabilizing the numerics. The regularization is therefore quite robust and we neglect the error for the benefit of a numerically stable expression for the polarization function. Without this subtraction, other approximations like the introduction of a cutoff are needed for the numerical calculation which usually lead to larger errors.

Furthermore, with color-superconducting condensates present, the polarization tensor $\Pi^{ab}_{\mu\nu}(k)$ \eq{eq:gpol} is not necessarily diagonal in color space. For 2SC or CFL-like phases, block structures arise for $(a,b)\in\lbrace 4,5 \rbrace$ and $(a,b)\in\lbrace 6,7 \rbrace$ with the structure
\beq
\begin{pmatrix}
\Pi^{44}_{\mu\nu}(k) & \Pi^{45}_{\mu\nu}(k)\\[1mm] 
\Pi^{54}_{\mu\nu}(k) & \Pi^{55}_{\mu\nu}(k)
\end{pmatrix}
\eeq
and the properties $\Pi^{44}_{\mu\nu}(k)=\Pi^{55}_{\mu\nu}(k)$ and $\Pi^{45}_{\mu\nu}(k)=-\Pi^{54}_{\mu\nu}(k)$.
These blocks can be diagonalized using the unitary matrix~\cite{Rischke:2000qz}
\beq
u=\frac{1}{\sqrt{2}}
\begin{pmatrix}
1 & -i\\
-i & 1
\end{pmatrix}.
\eeq
This allows us to use a color-diagonal gluon propagator $D^{aa}_{\mu\nu}(k)$ with complex components. 

\end{document}